\documentclass[a4paper,11pt]{article}
\usepackage{pos}
\usepackage{soul}
\usepackage{color}
\usepackage{mathtools}

\title{Doubly $\kappa$-deformed Yang models, Born - selfdual $\kappa$-deformed quantum phase spaces and two generalizations of Yang models}
\author*[a]{Jerzy Lukierski}
\author[b]{Anna Pacho\l} 

\affiliation[a]{Institute of Theoretical Physics, Wroc\l aw University,\\ pl. Maxa Borna 9, 50-205 Wroc\l aw, Poland}

\affiliation[b]{Department of Microsystems, University of South-Eastern Norway,\\ Campus Vestfold, Raveien 215, 3184 Borre, Norway}

\emailAdd{jerzy.lukierski@uwr.edu.pl}
\emailAdd{anna.pachol@usn.no}

\abstract{Recently it was shown that by using two different realizations of $\hat{o}(1,4)$ Lie algebra one can describe one-parameter standard Snyder model and two-parameter $\kappa$-deformed Snyder model. In this paper, by using the generalized Born duality and Jacobi identities we obtain from the $\kappa$-deformed Snyder model the doubly $\kappa$-deformed Yang model which provides the new class of quantum relativistic phase spaces. These phase spaces contain as subalgebras the $\kappa$-deformed Minkowski space-time as well as quantum $\tilde{\kappa}$-deformed fourmomenta and are depending on five independent parameters. Such a large class of quantum phase spaces can be described in $D=4$ by particular realizations of $\hat{o}(1,5)$ algebra, what illustrates the property that in noncommutative geometry different $D=4$ physical models may be described by various realizations of the same algebraic structure.
Finally, in the last Section we propose two new ways of generalizing Yang models: by introducing $\hat o(1,3+2N)$ algebras ($N=1,2\ldots$) we provide internal symmetries $O(N)$ symmetries in 
Kaluza-Klein extended Yang model, and by replacing the classical $\hat{o}(1,5)$ algebras which describe the algebraic structure of  Yang models by $\hat o(1,5)$ quantum groups with suitably chosen nonprimitive coproducts.}

\FullConference{
  Corfu Summer Institute 2023 "Workshop on Noncommutative and Generalized Geometry in String Theory, Gauge Theory and Related Physical Models"\\
  September 18 - September 25, 2023\\
  Corfu, Greece
}

\begin{document}
\maketitle

\tableofcontents

\section{Introduction}
At present it is still too early to know how the plausible final description of quantum gravity (QG) looks. However, there was found sufficient amount of indications that in order to describe QG one should modify, especially at the very short distances, the classical concepts of continuous space-time and the formulae defining the canonical quantum-mechanical phase space.
Since the discovery of quantum mechanics (QM), more than a hundred years ago, we understood that the quantization procedure implies the appearance of noncommutativity (NC) which is reflected e.g. in the limitations imposed on the simultaneous measurements of quantum-mechanical coordinates and momenta \footnote{In standard NC nonrelativistic QM the quantum-mechanical coordinates are described by the space coordinates and three-momenta. In the relativistic version of QM one introduces quantum-mechanical coordinates as Minkowski space-time coordinates and the Lorentz-covariant fourmomenta, with relativistic energy described by their fourth components.}.

In this paper we consider the general quantized framework, with quantum theory of fundamental interactions and QG both included. The historical development provided many arguments that QG, in the presence of matter described by quantum field theory (QFT) requires the introduction of NC quantum space-time and quantum noncanonical phase spaces. Usually, in QM one adds gravity only as the background field, but in QFT framework the gravitational degrees of freedom play also the dynamical role, what leads to the modification of canonical quantum-mechanical quantization rules.
In 1990s Dopplicher et al. \cite{Dop1, Dop2} have considered the modification of Heisenberg uncertainty relations in the presence of quantized gravitational degrees of freedom. It appeared that due to the interaction with the quantum dynamical gravitational background the quantum mechanical uncertainty relations are substantially modified. As a result, it has been shown that one cannot measure the ultra-short distances, which are smaller than the Planck length $\lambda_P\sim 10 ^{-33}\ \mathrm{cm}$, where
\begin{equation}\label{lp}
\lambda_P=\sqrt{\frac{\hbar G}{c^3}}\sim 1.02 \cdot 10 ^{-33}\ \mathrm{cm}.
\end{equation}
Because $\lambda_P$ depends on the Newton constant, i.e. the gravitational coupling constant, the appearance of absolute resolution in the measurements of minimal space-time distances indicates its gravitational origin
and leads to non-continuous structure of quantum space-time replacing the continuous classical space-time and the commutativity of classical coordinates. The NC structure of space-time in the presence of QG can be also linked  (see e.g. \cite{Bronstein} , \cite{Garay}) with the gravitational mechanism of creating microscopic black holes, which effectively, due to their quantum ($\hbar\neq 0 $) and gravitational ($G\neq 0$) nature (see \eqref{lp}), replaces the classical space-time geometry. These new geometric features lead us to the formalism of noncommutative geometry (NCG) which has been studied as a new tool of mathematical physics, with justified hopes for its applications to the description of QG effects \footnote{Unfortunately, from experimental side, the results of QG phenomenology till now are quite modest (see e.g. \cite{Addazi:2021xuf}). However, one should be optimistic if we observe the continuous progress in the construction of new measuring devices (e.g. located at the satellites) remarkably effective in the domain of astrophysics.}.  Natural application of NCG is the construction of the algebras describing NC quantum space-times and quantum phase spaces \footnote{{We interpret, in this lecture,  the words "quantum-deformed" or "$\kappa$-deformed" not rigorously;
% as according to the mathematical deformation theory, that the algebraic objects before and after deformation should be always nonisomorphic.
we permit the class of quantum phase spaces linked 
  by isomorphic maps since they may lead to different $D=4$ physical models.}}, in general case with NC coordinates and momenta.

The first relativistic NC $D=4$ models describing the algebras of quantum space-time coordinates were introduced in 1947 by Snyder \cite{Snyder}
and the algebras providing relativistic  quantum
phase spaces were proposed by C.N.Yang \cite{Yang}.
   
In Snyder model, by using $D=4$ dS algebra generators $\hat{M}_{ab}=(\hat{M}_{\mu\nu},\hat{M}_{4\mu})$ ($a,b=0,1,\ldots,4$), one introduces the following identification of NC space-time coordinates ($\mu,\nu=0,1,2,3$):
\begin{equation}
\hat{M}_{4\mu }=M \hat{x}_\mu\label{x_id}
\end{equation}
where $M$ is the inverse of the elementary length parameter $\lambda$ which plays the role of dimensionfull mass-like deformation parameter, frequently identified with the Planck mass $m_P$. The following set of algebraic relations describes the $D=4$ Snyder model \cite{Snyder}\footnote{Following the original formulation in ref.~\cite{Snyder, Yang} of Snyder and Yang models,  we will expose explicitly the dependence of algebraic formulas on the Planck constant. Such $\hbar$-dependent algebras provide the quantum-mechanical formulation of Snyder and Yang models.} with $\eta_{\mu\nu}=diag(-1,1,1,1)$:
\begin{equation}
\lbrack \hat{x}_{\mu },\hat{x}_{\nu }]=\frac{i\hbar}{M^2}\hat{M}_{\mu
\nu },  \label{snyderx}
\end{equation}
\begin{equation}
\lbrack \hat{M}_{\mu \nu },\hat{x}_{\rho }]=i\hbar (\eta _{\mu \rho }\hat{x}%
_{\nu }-\eta _{\nu \rho }\hat{x}_{\mu }), \label{snyderMx}
\end{equation}
\begin{equation}
\lbrack \hat{M}_{\mu \nu },\hat{M}_{\rho \tau }]=i\hbar (\eta _{\mu \rho
}\hat{M}_{\nu \tau }-\eta _{\mu \tau }\hat{M}_{\nu \rho }+\eta _{\nu \tau }\hat{M}_{\mu \rho
}-\eta _{\nu \rho }\hat{M}_{\mu \tau }) \label{snyderMM}
\end{equation}
where relations (\ref{snyderMx}) express the Lorentz covariance of Snyder model and (\ref{snyderMM}) describes the Lorentz algebra which provides the extension of quantum mechanical nonrelativistic angular momentum $\hat o(3)$ algebra, with $\hbar$-dependence used as in standard books on Quantum Mechanics (see e.g.\cite{Dirac,Landau}).

In the present paper we will describe the class of NCG models which define new family of noncanonical quantum relativistic phase spaces (see also \cite{PLB2024}). 
The plan of our paper is the following:

In Sect. 2 we recall the construction of $\kappa$-deformed Snyder model and its description by particular realization of $\hat{o}(1,4)$ algebra \cite{MM_PLB814,MM_PRD104,MP_2021,LMMP_PLB838}. In Sect. 3 we describe the standard Yang model \cite{Yang} which is characterized by the self-duality under the Born map\footnote{We put $\hbar=c=1$, i.e. $[M]=L^{-1}$. The Born map relates the coordinates and parameters with opposite length or mass dimensions, the dimensionless variables (e.g. rotation generators) are Born self-dual.} %known already since 1938 
\cite{Born1, Born2}
\begin{equation}\label{born}
B:\quad \hat{x}_\mu\rightarrow \hat{q}_\mu,\quad \hat{q}_\mu\rightarrow - \hat{x}_\mu, \quad\hat{M}_{\mu\nu}\leftrightarrow \hat{M}_{\mu\nu},\quad M=\frac{1}{\lambda}\leftrightarrow R
\end{equation}
where ($\hat{x}_\mu,\hat{q}_\mu$) are the NC coordinates of quantum relativistic phase space and two parameters ($M, R$, $[M]=L^{-1}, [R]=L$)\footnote{One often identifies $M$ with Planck mass $m_P$, $\lambda_P$ with Planck length $\lambda_P$, and $R$ with de Sitter (dS) radius of the Universe.} introduce the constant Riemannian curvatures in space-time and fourmomenta sectors of the Yang model.
We introduce, in Sect. 4, doubly $\kappa$-deformed Yang model with the pair of algebraic sectors describing $\kappa$-deformed quantum space-times and quantum momenta.
We perform the following two steps:
\begin{enumerate}
\item In order to describe the $\kappa$-deformations along time-like/light-cone/space-like directions in $D=4$ Minkowski space $M^{1,3}$ and in the four dimensional space $P^{1,3}$ of relativistic fourmomenta, one should introduce two triplets of constant suitably normalized  fourvectors:
\begin{itemize}
\item[i)] in $M^{1,3}$ the fourvector $a_\mu$ with three possible invariant lengths $a^2 = a_\mu a^\mu=(1,0,-1)$;
\item[ii)] in $P^{1,3}$ the fourvector $b_\mu$ with analogous normalized three lengths $b^2= b_\mu b^\mu=(1,0,-1)$.
The presence of fourvectors $a_\mu,\ b_\mu$ permits to describe double $\kappa$-deformed Yang models in relativistic-covariant way.
\end{itemize}
Further, we introduce (besides the parameters $M, R$ characterizing standard Yang model) the pair of mass-like deformation parameters $\kappa$, $\tilde{\kappa}$ ($[\kappa]=[\tilde{\kappa}]=L^{-1}$) describing respectively $\kappa$-deformations in space-time and fourmomenta sectors and additional dimensionless parameter $\rho ([\rho]=L^0)$ which is permitted by the Jacobi identities.
\item We introduce the following enlargement of the Born map \eqref{born} (see also \cite{PLB2024})
\begin{equation}\label{born_ext}
\tilde{B}:\quad \kappa\leftrightarrow \frac{1}{\tilde{\kappa}},\quad a_\mu\rightarrow b_\mu,\quad b_\mu\rightarrow -a_\mu.
\end{equation}
In order to obtain double $\kappa$-deformed Yang models we apply the generalized Born map $\mathcal{B}=B\oplus \tilde{B}$ \eqref{born} and  \eqref{born_ext} to the $\kappa$-deformed Snyder model.
% introduced, \ap{by Meljanac et al.} \cite{kappa-Snyder}, \cite{kappa-Snyder1} (see also \cite{MM_PLB814},\cite{MM_PRD104}, \cite{MP_2021}) in 2020.
The generalized Born map  permits to obtain from $\kappa$-deformed Snyder model the terms in the algebraic formulation of our new extended Yang model. It appears however (see Sect. 5) that by using the Jacobi identities one can add in algebraically consistent way additional term which is self-dual under the generalized Born map $\mathcal{B}$ and dimensionless fifth parameter $\rho$ ($ [\rho]=L^0$) besides $M, R, \kappa$ and $\tilde{\kappa}$. One of such terms, proportional to $\rho$, occurs in Triply Special Relativity (TSR) model \cite{TSR}. We add that in Sect. 5, we consider
two different $\kappa$-deformations, described by two independent mass-like parameters $\kappa, \tilde{\kappa}$. The first providing the well known $\kappa$-deformation of quantum space-time and the second introducing new $\tilde{\kappa}$-deformation in NC four-momentum space.
\end{enumerate}
Further in Sect. 6 we point out that two Yang type models: standard one, and doubly $\kappa$-deformed, can be described by two different realizations of $\hat{o}(1,5)$ Lie algebra, i.e. both models provide different NC physical models which are described by the same underlying algebraic structure.

Finally, Sect. 7 contains outlook and final remarks. In particular we comment how Yang models can be generalized in order to describe the internal symmetry multiplets of quantum space-times and quantum momenta fourvectors, what is the subject of our very recent studies \cite{27}.

\section{$\kappa$-deformed Snyder models and particular realizations of $D=4$ de-Sitter algebra $\hat o(1,4)$}

The standard $\kappa$-deformed Minkowski space-time $\hat{x}_\mu$ ($\mu=0,1,2,3$)
can be described by the classical three-dimensional space coordinates $\hat x_\mu=x_i$ ($i=1,2,3$) and nonclassical quantum time variable $\hat{x}_0$ which satisfies the following well-known commutation relations:
\begin{equation}\label{kappaM}
\lbrack \hat{x}_{0},\hat{x}_{i}]=\frac{i\hbar}{\kappa}\hat{x}_i,\qquad \lbrack \hat{x}_{i},\hat{x}_{j}]=0.
\end{equation}
If we introduce the fourvector $a_\mu=(1,0,0,0)$ (i.e. time-like case) one can describe the above relations \eqref{kappaM} in the following form \footnote{The formula \eqref{kappaMa} appeared firstly in \cite{kMa}.}:
\begin{equation}\label{kappaMa}
\lbrack \hat{x}_{\mu},\hat{x}_{\nu}]=\frac{i\hbar}{\kappa}(a_\mu\hat{x}_\nu -a_\nu \hat{x}_{\mu})
\end{equation}
and then generalize these to light-like and space-like cases, depending on the choice of the fourvector $a_\mu$. In this way the choice of constant dimensionless four-vector $a_\mu$ determines three types of the $\kappa$-deformations of quantum Minkowski space-times: time-like (or standard one) if $a_\mu a^\mu=-1$ (we use the metric $\eta_{\mu\nu}=(-1,1,1,1)$), tachyonic if $a_\mu a^\mu =1$ and light-like if $a_\mu a^\mu=0$.

Extending the right hand side of \eqref{kappaMa} by the "Snyder term" (cf. \eqref{snyderx}) $\frac{i\hbar}{M^2}M_{\mu\nu}$ \footnote{Alternatively, one can say that we extend the Snyder relations \eqref{snyderx}-\eqref{snyderMx} by the $\kappa$-Minkowski space-time \eqref{kappaM}.} 
we obtain the algebra of $\kappa$-deformed Snyder model \cite{kappa-Snyder,kappa-Snyder1}, unifying $\kappa$-Minkowski and quantum Snyder space-time:
\begin{equation} 
\lbrack \hat{x}_{\mu },\hat{x}_{\nu }]=i\hbar\left[\frac{1}{M^2}\hat{M}_{\mu
\nu}+\frac{1}{\kappa}(a_\mu \hat{x}_\nu - a_\nu \hat{x}_\mu)\right].  \label{snyderkx}
\end{equation}
If we put $M\to\infty$ in (\ref{snyderkx}) we obtain the generalized $a_\mu$-dependent $\kappa$-deformed Minkowski space-time, with $\hat{x}=a^\mu \hat{x}_\mu$
describing the unique NC quantum coordinate; if $\kappa\to\infty$ we obtain the standard Snyder model.

The relativistic covariance relation of $\kappa$-deformed Snyder model looks as then follows (compare with \eqref{snyderMx}):
\begin{equation}
\lbrack \hat{M}_{\mu \nu },\hat{x}_{\rho }]=i\hbar \left[\eta _{\mu \rho }\hat{x}_{\nu }-\eta _{\nu \rho }\hat{x}_{\mu }
+\frac{1}{\kappa}(a_\mu \hat{M}_{\rho\nu} - a_\nu \hat{M}_{\rho\mu})\right].  \label{snyderMkx}
\end{equation}

The formulae \eqref{snyderkx}-\eqref{snyderMkx} can be described as Lie-algebraic set of relations of the algebra $\hat{o}(1,4;g)$ with the metric $g\equiv g_{ab}$, $a,b=0,1,2,3,4$ (see \cite{MM_PLB814},\cite{MM_PRD104}).
\begin{equation}
\lbrack \hat{X}_{a b },\hat{X}_{c d }]=i\hbar (g_{a c }
\hat{X}_{b d }-g_{b c }\hat{X}_{a d }+g_{b d }\hat{X}_{a c }-g_{a d }\hat{X}_{b c }).  \label{alg_sog}
\end{equation}
where $\hat{X}_{\mu\nu}=\hat{M}_{\mu\nu}$ are $D=4$ Lorentz generators satisfying \eqref{snyderMM} and $\hat{X}_{\mu 4}=-\hat{X}_{\mu 4}=M\hat{x}_\mu$. 

The metric $g_{ab }$ is given by the following $5\times 5$ matrix:
\begin{equation}
g=\left(
\begin{array}{ccccc}
-1 & 0 & 0 & 0 & \frac{M}{\kappa}a_0 \\
0 & 1 & 0 & 0 & \frac{M}{\kappa}a_{1} \\
0 & 0 & 1 & 0 & \frac{M}{\kappa}a_2 \\
0 & 0 & 0 & 1 & \frac{M}{\kappa}a_3 \\
\frac{M}{\kappa}a_{0} & \frac{M}{\kappa}a_{1} & \frac{M}{\kappa}a_2 & \frac{M}{\kappa}a_{3} & 1
\end{array}
\right). 
\end{equation}
We see that $g^T=g$ and one can calculate that $\det g=\frac{M^2}{\kappa^2}a_\mu a^\mu -1$.

Only for tachyonic $\kappa$-deformation and when $M=\kappa$ one obtains that $\det g=0$. If $\det g\neq 0$ one can introduce the pseudoorthogonal matrix $O_{ab}$ satisfying the relation
\begin{equation}
g_{ab}=\left( O\eta O^{T}\right) _{ab}.
\end{equation}
One can choose the following triangular matrix
\begin{equation}
O=\left(
\begin{array}{ccccc}
1 & 0 & 0 & 0 & 0 \\
0 & 1 & 0 & 0 & 0 \\
0 & 0 & 1 & 0 & 0 \\
0 & 0 & 0 & 1 & 0 \\
-\frac{M}{\kappa}a_{0} & \frac{M}{\kappa}a_{1} & \frac{M}{\kappa}a_2 & \frac{M}{\kappa}a_{3} & d
\end{array}
\label{macierz}
\right)
\end{equation}
where $d =\sqrt{-\det g}=(1-\frac{M^2}{\kappa^2}a_\mu a^\mu)^\frac{1}{2}$ and one can introduce the new basis:
\begin{equation}
\hat{X}_{ab }=(O\hat{x}O^T)_{ab}=(\hat{X}_{\mu\nu}=\hat{M}_{\mu\nu}, \hat{X}_{\mu 4}=M\hat{x}_\mu).
\label{basis}
\end{equation}
One can check that due to relations \eqref{alg_sog} and \eqref{basis} the generators $\hat{x}_{ab}=(\hat{x}_{\mu\nu}=\hat{M}_{\mu\nu}, \hat{x}_{\mu 4}=M\hat{x}_\mu)$ describe the Lie algebra $\hat{o}(1,4;\eta_{\mu\nu})\equiv \hat{o}(1,4)$:
\begin{equation}
\lbrack \hat{x}_{a b },\hat{x}_{c d }]=i\hbar (\eta_{a c }
\hat{x}_{b d }-\eta_{b c }\hat{x}_{a d }+\eta_{b d }\hat{x}_{a c }-\eta_{a d }\hat{x}_{b c }).  \label{alg_so}
\end{equation}
where $\eta_{ab}=(\eta_{\mu\nu},1)$.
The explicit formulae relating the bases $\hat{X}_{ab}$ and $\hat{x}_{ab}$ look as follows (see \cite{MM_PRD104},\cite{LMMP_PLB838})
\begin{equation}  \label{Xxrel1}
\hat{X}_{\mu\nu}=\hat{x}_{\mu\nu}=\hat{M}_{\mu\nu}
\end{equation}
\begin{equation}\label{Xxrel2}
\hat{X}_{\mu}=d \hat{x}_\mu+\frac{1}{\kappa}\hat{M}_{\mu\nu}a^\nu
\end{equation}
or equivalently\footnote{Compare with the change of basis between Snyder model and $\kappa$-Minkowski space-time, for the choice $a_\mu=(1,0,0,0)$, which were considered in \cite{Kowalski-Glikman:2003qjp}, \cite{ABAP_PRD2010}, \cite{Borowiec:2010yw}.}
\begin{equation}\label{Xxrel3}
\hat{x}_\mu=\frac{1}{d}(\hat{X}_{\mu}-\frac{1}{\kappa}\hat{M}_{\mu\nu}a^\nu).
\end{equation}
This set of formulae \eqref{Xxrel1}-\eqref{Xxrel3} implies in the limit $M\to\infty$ (in particular if $a_\mu a^\mu=-1$, $d$ is real and $d\to\infty$) that $\hat{x}_\mu$ satisfying $\hat{o}(1,4)$ relations \eqref{alg_so} vanishes, i.e. one cannot describe $\kappa$-deformed Minkowski space-time as the realization of $\hat{o}(1,4)$ algebra.
In other limit, when $\kappa\to\infty$ we get $d=1$ and $\hat{x}_\mu$ describing Snyder space-time \eqref{snyderx}-\eqref{snyderMM}. 

If $M$ is finite and $\kappa\neq 0$, the algebra of $\kappa$-deformed Snyder model \eqref{snyderkx} can be described by  
various realizations of $\hat{o}(1,4)$ linked by the suitably chosen matrices $O_{ab}$ \eqref{macierz}.

\section{Yang models as Born-dual extensions of Snyder models}

In order to obtain Yang model we add to the relation \eqref{x_id} analogous geometric formula for NC curved fourmomenta $\hat{q}_\mu$:
\begin{equation}
\hat{M}_{5\mu }=R \hat{q}_\mu\label{q_id}
\end{equation}
satisfying the commutation relation
\begin{equation}
\lbrack \hat{q}_{\mu },\hat{q}_{\nu }]=\frac{i\hbar}{R^2}\hat{M}_{\mu
\nu },  \label{yangq}
\end{equation}
where $[R]=L$ describes the length parameter, which in astrophysical applications was often identified with the cosmological $D=4$ dS radius of the Universe. In order to obtain the set of $\hat{o}(1,5)$ Lie algebra generators (with indices $A, B=0, 1, \ldots, 5$; {$a, b=0, 1, \ldots, 4$}):
\begin{equation}\label{MAB}
  \hat{M}_{AB}=(\hat{M}_{ab},  \hat{M}_{a5})= (\hat{M}_{\mu\nu},  \hat{M}_{4\mu},  \hat{M}_{5\mu},  \hat{M}_{45})
\end{equation}
we define new scalar generator $\hat{r}$ by the relation
\begin{equation}\label{M45}
 \hat{M}_{45}=MR\hat{r}.
\end{equation}
In $D=4$ space-time the generator $\hat{r}$ describes the quantum $\hat{o}(2)$ symmetry which acts on the doublet $(\hat{x}_\mu,\hat{q}_\mu)$
as follows\footnote{If we introduce the $\hat{o}(2)$ rotations generated by $\hat{r}$:
\begin{eqnarray}\label{change}
\hat{x}'_\mu&=&\cos\varphi \hat{x}_\mu+\sin\varphi \hat{q}_\mu\nonumber\\
\hat{q}'_\mu&=&-\sin\varphi \hat{x}_\mu+\cos\varphi \hat{q}_\mu
\end{eqnarray}
and choose $\varphi=\pi$ one obtains from \eqref{change} the Born map \eqref{born}. We see that the continuous map \eqref{change} describes the extension of discrete Born map \eqref{born} defined by \eqref{change} if $\varphi=\frac{\pi}{2}$.
%\ap{or for other choices of the angle $\varphi$?}.
}
\begin{equation}\label{rxrq}
[\hat{r},\hat{x}_\mu]=\frac{i\hbar}{M^2} \hat{q}_\mu, \qquad [\hat{r},\hat{q}_\mu]=-
\frac{i\hbar}{R^2} \hat{x}_\mu.
\end{equation}
The remaining $\hat{o}(1,5)$ algebra relations satisfied by generators \eqref{MAB} are the following:
\begin{itemize}
\item[i)] relativistic generalized Heisenberg relation:
\begin{equation}\label{xq}
[\hat{x}_\mu, \hat{q}_\nu]=i\hbar\eta_{\mu\nu}\hat{r},
\end{equation}
where the case $\hat{r}=\mathrm{1}$ occurring in canonical commutation relations is replaced by the unique scalar $\hat{r}$, i.e. 
\begin{equation}\label{Mr}
 [\hat{M}_{\mu\nu},\hat{r}]=0.
\end{equation}
\item[ii)] relativistic covariance relations:
\begin{equation}
\lbrack \hat{M}_{\mu \nu },\hat{x}_{\rho }]=i\hbar (\eta _{\mu \rho }\hat{x}_{\nu }-\eta _{\nu \rho }\hat{x}_{\mu }).  \label{yangMkx}
\end{equation}
\begin{equation}
\lbrack \hat{M}_{\mu \nu },\hat{q}_{\rho }]=i\hbar (\eta _{\mu \rho }\hat{q}_{\nu }-\eta _{\nu \rho }\hat{q}_{\mu }).  \label{yangMkq}
\end{equation}
It can be checked that the generators \eqref{MAB} satisfy the $\hat{o}(1,4)$ Lie algebra relations
\begin{equation}
\lbrack \hat{M}_{AB},\hat{M}_{CD}]=i\hbar (\eta _{AC}\hat{M}_{BD}-\eta _{AD}\hat{M}_{BC}+\eta _{BD}\hat{M}_{AC}-\eta _{BC }\hat{M}_{AD}) \label{MM}
\end{equation}
where $\eta_{AB}=diag(-1,1,\ldots,1)$.
\end{itemize}
As it was already mentioned in Sect. 1, the Yang algebraic relations \eqref{snyderx}-\eqref{snyderMM},  \eqref{yangq}, \eqref{rxrq}-\eqref{MM} are self-dual under the Born map relations \eqref{born} extended by Born self-duality relation of $\hat{r}$:
\begin{equation}
B:\quad \hat{r}\leftrightarrow \hat{r}
\end{equation}
in analogy with the Lorentz generators in \eqref{born}.

We did show that Yang algebra can be obtained as the Born extension of Snyder algebra and describes the geometric model of quantum-deformed Heisenberg algebra with NC space-time and fourmomenta coordinates and $\hat{o}(2)$ internal symmetries generator. In the next section we will introduce the Born extension of $\kappa$-deformed Snyder model, what required the enlargement of the map \eqref{born} and leads to the Born-dual $\kappa$-deformation of fourmomenta sector.
\section{Generalized Born map and doubly kappa-deformed Yang models as Born-self-dual extensions of kappa-deformed Snyder models}
By performing generalized Born map $\mathcal{B}=B\oplus \tilde{B}$ (see \eqref{born} and  \eqref{born_ext}) on the $\kappa$-deformed Snyder algebra relations \eqref{snyderMM}, \eqref{snyderkx}, \eqref{snyderMkx}, we obtain doubly $\kappa$-deformed Yang model described by the following relations:
\begin{equation}\label{qqk}
[\hat{q}_\mu,\hat{q}_\nu]=i\hbar \left[\frac{\hat{M}_{\mu\nu}}{R^2}+\tilde{\kappa}(b_\mu\hat{q}_\nu-b_\nu\hat{q}_\mu)\right].
\end{equation}
\begin{equation}
[ \hat{M}_{\mu\nu},\hat{q}_{\rho}]=i\hbar \left[\eta _{\mu\rho}\hat{q}_{\nu}-\eta
_{\nu\rho}\hat{q}_{\mu}+\tilde{\kappa}(b_\mu\hat{M}_{\rho\nu}-b_\nu\hat{M_{\rho\mu}})\right].\label{Mqkappa}
\end{equation}
These relations depend on a new set of fourvectors $b_\mu$ and supplementary mass-like parameter $\tilde{\kappa}$, both characterizing new type of $\tilde{\kappa}$ deformation in fourmomentum space. 

In Yang model, besides the NC space-time coordinates \eqref{snyderkx}, do appear the NC fourmomenta $\hat{q}_\mu$ \eqref{qqk}, what permits to consider the new algebra of quantum  phase space coordinates $(\hat{x}_\mu,\hat{q}_\mu)$. The generalized Heisenberg algebra is obtained if we calculate the Jacobi identities for the set of operators $(\hat{x}_\mu,\hat{q}_\mu)$ and $(\hat{r},\hat{M}_{\mu\nu})$.
We obtain the following algebraically consistent generalization of the relations \eqref{rxrq}, \eqref{xq}, \eqref{Mr} characterizing standard Yang model

\begin{equation}\label{xqd}
[\hat{x}_\mu,\hat{q}_\nu]=i\hbar\left(\eta_{\mu\nu}\hat{r}+\tilde{\kappa}b_\mu\hat{x}_\nu -\frac{a_\nu}{\kappa} \hat{q}_\mu +\frac{\rho}{MR}\hat{ M}_{\mu\nu}\right),
\end{equation}
\begin{equation}\label{rxd}
[\hat{r},\hat{x}_\mu]=i\hbar\left(\frac{1}{M^2}\hat{q}_\mu-\frac{1}{MR}\rho\hat{x}_\mu - \frac{a_\mu}{\kappa}\hat{r}\right),
\end{equation}
\begin{equation}\label{rqd}
[\hat{r},\hat{q}_\mu]=i\hbar\left(-\frac{1}{R^2}\hat{x}_\mu+\frac{1}{MR}\rho\hat{q}_\mu - \tilde{\kappa} b_\mu\hat{r}\right),
\end{equation}
\begin{equation}
\lbrack \hat{r}, \hat{M}_{\mu \nu }]=-i\hbar \left[\frac{1}{\kappa}(a_\mu\hat{q}_{\nu }-
a_\nu\hat{q}_\mu)-\tilde{\kappa}(b_\mu
\hat{x}_{\nu } - b_\nu \hat{x}_{\mu})\right].  \label{rMd}
\end{equation}
Interestingly, if we pass from $\kappa$-deformed Snyder to doubly $\kappa$-deformed Yang models, we observe that
\begin{equation}
\lbrack \hat{r}, \hat{M}_{\mu \nu }]=0\quad\xrightarrow[\hat{q}_\mu,\tilde{\kappa} ]{\text{added}}\quad \lbrack \hat{r}, \hat{M}_{\mu \nu }]\neq 0
\end{equation}
i.e. in doubly $\kappa$-deformed Yang model the commutator between Lorentz and internal symmetries generators is not vanishing \footnote{Such noncommutativity implies that the Coleman-Mandula theorem is not valid for doubly $\kappa$-deformed Yang model.}. Other property which follows from studying the general solutions of Jacobi identities for the set of 15 generators $(\hat{x}_\mu, \hat{M}_{\mu\nu}, \hat{q}_\mu,\hat{r})$ leads to the appearance in formulae \eqref{xqd}-\eqref{rqd} of the dimensionless parameter $\rho$ ($[\rho]=L^0$) which in the relation \eqref{xqd} describing quantum  relativistic phase space algebra is multiplied by Lorentz generator $\hat{M}_{\mu\nu}$.
Further, from relations \eqref{xqd}-\eqref{rqd}, it follows that nonvanishing parameter $\rho$ appears only if both parameters $M$ and $R$ are finite, i.e. both quantum space-time and quantum fourmomenta space are curved. This property agrees with the features of TSR model \cite{TSR} where both of these parameters are finite, nonvanishing and proportional to $\rho.$
%
%$\xrightarrow[\text{world}]{\text{hello}}$
%$\xrightarrow[g(x)]{f(x)}$

%\section{Kappa-deformations of quantum fourmomenta sector}

\section{Doubly kappa-deformed Yang models as described by particular realizations of $\hat o(1,5)$ Lie algebra}

In Sect. 2 we have shown that the $\kappa$-deformed Snyder model can be described by particular class of realizations $\hat{o}(1,4)$ algebra. In Sect. 3 we recalled (see \eqref{MAB})
that standard Yang model \cite{Yang} can be algebraically described by $\hat{o}(1,5)$ algebra 
\eqref{MM}. Further in Sect. 4 we introduced $\kappa$-deformed Yang model (see also \cite{LMMP_PLB838}) which will be presented as a particular realization of $\hat{o}(1,5)$ algebra.

We will show that by the generalization of $\kappa$-deformed Snyder model \cite{MM_PLB814}-\cite{LMMP_PLB838}, by Lie algebra $\hat{1,4;g}$ (Sect. 2) one can describe doubly $\kappa$-deformed Yang model by Lie algebra $\hat{o}(1,5;g^{(Y)})$
 \begin{equation}
\lbrack \hat{M}^{(Y)}_{AB},\hat{M}^{(Y)}_{CD}]=i\hbar (g^{(Y)}_{AC}\hat{M}^{(Y)}_{BD}-g^{(Y)}_{AD}\hat{M}^{(Y)}_{BC}+g^{(Y)}_{BD}\hat{M}^{(Y)}_{AC}-g^{(Y)}_{BC }\hat{M}^{(Y)}_{AD}). \label{Yang_gMM}
\end{equation}
The dimensionless symmetric metric components $g^{(Y)}_{AB}$ with the signature $(-1,1,\ldots,1)$
describe the doubly $\kappa$-deformed Yang models which depend on five deformation parameters $(M,R,\kappa,\tilde{\kappa},\rho)$  $([M]=L^{-1},[R]=L,[\kappa]=[\tilde{\kappa}]=L^{-1},[\rho]=L^0)$ and the pair of constant dimensionless four-vectors $a_A=(a_\mu,0,0), b_A=(b_\mu,0,0)$, $\mu=0,1,2,3$ which besides the known $\kappa$-deformations in quantum $D=4$ space-time introduce the Born dual $\tilde{\kappa}$-deformations of $D=4$ quantum fourmomenta sector of Yang algebra. 
The metric $g^{(Y)}_{AB}$ is determined by the following assignments of the generators (see (\ref{x_id}), (\ref{q_id}) and \eqref{M45}):
\begin{equation}
M^{(Y)}_{AB}=\left(\hat{M}_{\mu\nu},\ \hat{M}^{(Y)}_{4\mu}=M\hat{x}_\mu,\ \hat{M}^{(Y)}_{5\mu}=R\hat{q}_\mu,\ \hat{M}^{(Y)}_{45}=MR\hat{r}\ \right)\label{M_Y_AB}
\end{equation}
where $[M^{(Y)}_{AB}]=L^0$ (dimensionless).
$\hat{M}_{\mu\nu}$ describes $D=4$ Lorentz algebra and the scalar $\hat{r}$ is the generator of the $\hat{o}(2)$ internal symmetries, as before.
Relations (\ref{Yang_gMM}) describe the doubly $\kappa$-deformed Yang model if we consider the following components of the $D=6$ metric tensor:
\begin{equation}\label{g_matrix}
g_{AB}^{\left( Y\right) }=\left(
\begin{array}{ccc}
\eta _{\mu \nu } & g_{\mu 4}^{\left( Y\right) } & g_{\mu 5}^{\left( Y\right)
} \\
g_{4\nu }^{\left( Y\right) } & g_{44}^{\left( Y\right) } & g_{45}^{\left(
Y\right) } \\
g_{5\nu }^{\left( Y\right) } & g_{54}^{\left( Y\right) } & g_{55}^{\left(
Y\right) }%
\end{array}%
\right)
\end{equation}
where\footnote{{We recall that one can choose three types of constant four-vectors} $a_{\mu
}$ and $b_{\mu }$, with Lorentz-invariant lengths $\left( -1,0,1\right) $,
which select three types of $\kappa $ and $\tilde{\kappa}$-deformations: time-like
(or standard one), light-like and tachyonic. The
four-vectors $a_{\mu }$ and $b_{\mu }$ determine the quantum NC phase space
components $a^{\mu }\hat{x}_{\mu }$ and $b^{\mu }\hat{p}_{\mu }$, which due
to the double $\kappa -$deformations $\left( \kappa \neq 0,\tilde{\kappa}%
\neq 0\right) $ break explicitly the Lorentz covariance (compare with (\ref{snyderMkx}) and (\ref{Mqkappa})).}
\begin{equation}\label{g_coeff}
g_{\mu 4}^{\left( Y\right) }=g_{4\mu }^{\left( Y\right) }=\frac{M}{%
\kappa }a_\mu,\,\quad g_{\mu
5}^{\left( Y\right) }=g_{5\mu }^{\left( Y\right) }=R\tilde{\kappa}b_{\mu
},\quad g_{45}^{\left( Y\right) }=g_{54}^{\left( Y\right) }=\rho ,\quad
g_{44}^{\left( Y\right) }=g_{55}^{\left( Y\right) }=1
\end{equation}
with the pair of length parameters ($\lambda =M^{-1},R$) (or equivalently the pair of mass parameters $(M=\lambda^{-1}, \tilde M=R^{-1})$),
 the mass-like parameters $\left( \kappa ,\tilde{\kappa}\right) $ and the dimensionless parameter $\rho $ what implies that $g^{(Y)}_{AB}$ are dimensionless ($[g^{(Y)}_{AB}]=L^0$) in consistency with relations (\ref{Yang_gMM}).

The algebra (\ref{Yang_gMM}) for any choice of symmetric metric $g^{(Y)}_{AB}$ satisfies two
important properties:
\smallskip

i) By direct calculation one can show that the Lie algebra (\ref{Yang_gMM}) satisfies
Jacobi identities.

ii) For any nondegenerate symmetric metric $g^{(Y)}_{AB}$ with the signature described
by diagonal matrix $\eta _{AB}$ one can find $\left( 6\times 6\right)$-dimensional linear map $\mathbb{S}=S_{AB}$ satisfying the relation
\begin{equation}\label{gtrans}
\mathbf{g}^{\left( Y\right) }=\mathbb{S}\mathbb{\eta} \mathbb{S}^{T},\qquad \mathbf{g}^{\left( Y\right) }\equiv
\mathbf{g}_{AB}^{\left( Y\right) },\qquad \mathbb{\eta}=\eta_{AB}.
\end{equation}
Then one can relate the Lie algebras (\ref{YangMM}) and (\ref{Yang_gMM}) by the following maps
\begin{equation}\label{Mtrans}
\hat{M}^{(Y)}_{AB}=(\mathbb{S}\,\hat{\mathbb{M}}\mathbb{S}^T)_{AB} \qquad\longleftrightarrow\qquad \hat{M}_{AB}=(\mathbb{S}^{-1}\hat{\mathbb{M}}^{(Y)}(\mathbb{S}^T)^{-1})_{AB}.
\end{equation}
It follows that the matrix $\mathbb{S}$ satisfying relations (\ref{gtrans},\ref{Mtrans}) is not unique, with arbitrariness described by the pseudoorthogonal $6x6$ matrix $\mathbb{O}$, where $\mathbb{O}\eta\mathbb{O}^T=\eta.$ For concrete choice (\ref{gtrans}) of the matrix $\mathbf{g}^{(Y)}$ we choose %the symmetric
$6\times 6$ matrix $\mathbb{S}$ parametrized as follows\footnote{For the simplicity of formulae in (\ref{esse}), we introduce the shorthand notation
$$
g_{\mu 4}^{\left( Y\right) }=g_{4\mu }^{\left( Y\right) }=g_{\mu },\quad
g_{\mu 5}^{\left( Y\right) }=g_{5\mu }^{\left( Y\right) }=h_{\mu },
$$
$$
g_{44}^{\left( Y\right) }=g_{4},\quad g_{45}^{\left( Y\right)
}=g_{54}^{\left( Y\right) }=\rho ,\quad g_{55}^{\left( Y\right) }=h_{5}.
$$
}
\begin{equation}\label{esse}
\mathbb{S}=\left(
\begin{array}{cccccc}
-1 & 0 & 0 & 0 & 0 & 0 \\
0 & 1 & 0 & 0 & 0 & 0 \\
0 & 0 & 1 & 0 & 0 & 0 \\
0 & 0 & 0 & 1 & 0 & 0 \\
g_{0} & g_{1} & g_{2} & g_{3} & a & d \\
h_{0} & h_{1} & h_{2} & h_{3} & c & b%
\end{array}%
\right) \end{equation}
with parameters $a$, $b$, $c$, $d$ satisfying the conditions
%(note $g_ih_i\id-g_0h_0+g_1h_1+g_2h_2+g_3h_3$)
\begin{eqnarray}\label{25a}
&&a^2+d^2=g_4-g_\mu g^\mu=A,\cr
&&b^2+c^2=h_5-h_\mu h^\mu=B,\cr
&&ac+bd=\rho-g_\mu h^\mu=C.
\end{eqnarray}
We can pass to lower triangular $\mathbb{S}$ matrix if we put $d=0$ in the formulae (\ref{esse},\ref{25a}). In such a case 
%Assuming ${A+B-2\epsilon'\Delta}\ge0$, 
the set of equations (\ref{25a}) has the following solutions:
\begin{equation}
a=\epsilon\sqrt{A},\qquad b=\epsilon'\sqrt{B-\frac{C^2}{A}},\qquad c=\epsilon\frac{C}{\sqrt{A}},
\end{equation}
with $\epsilon,\epsilon'=\pm1$.
While considering the case of $d\neq 0$ we obtain:
\begin{eqnarray}
a &=&\epsilon \sqrt{A-d^{2}} \\
b &=&\frac{C}{A}d+\epsilon \left( \sqrt{\frac{AB-C^{2}}{A-d^{2}}}%
-\allowbreak \frac{d^{2}}{A}\sqrt{\frac{AB-C^{2}}{A-d^{2}}}\right)  \\
c &=&\epsilon \allowbreak \frac{\sqrt{A-d^{2}}}{A}\left( C-d\epsilon \sqrt{%
\frac{AB-C^{2}}{A-d^{2}}}\right).
\end{eqnarray}

In doubly $\kappa$-deformed Yang model one can select nine classes of double $\kappa$-deformations depending on the Lorentz-covariant normalized length values of the dimensionless fourvectors $a_\mu, b_\mu$ which are related with the fourvectors $g_\mu$ and $h_\mu$ from the matrix (\ref{esse}) if we choose
\begin{equation}
g_\mu=\frac{M}{\kappa}a_\mu \rightarrow A=1-\frac{M^2}{\kappa^2}a_\mu a^\mu,
\end{equation}
\begin{equation}
{h}_\mu=R\tilde{\kappa}b_\mu \rightarrow B=1-R^2\tilde{\kappa^2}b_\mu b^\mu
\end{equation}
and 
\begin{equation}
C=\rho-\frac{M}{\kappa}R\tilde{\kappa}a_\mu b^\mu.
\end{equation}
The nine classes of double $\kappa$-deformations are obtained by the choices of the parameters $\tilde{\epsilon},  \tilde{\epsilon}'=(\pm 1,0)$, where $a_\mu a^\mu =\tilde{\epsilon}$ and
$b_\mu b^\mu =\tilde{\epsilon}'$.

\section{Outlook}
In this paper we presented a new class of quantum phase spaces described by doubly $\kappa$-deformed Yang models, with noncomutative both space-time and fourmomenta  coordinates (see also \cite{PLB2024}).

It should be observed that the
large class of recently considered relativistic quantum phase spaces deals with NC quantum space-times, but keeps the momenta commutative (see e.g. \cite{24,25}). Such relativistic models were obtained as various generalizations of Snyder algebra \cite{Snyder}, see \eqref{snyderx}-\eqref{snyderMM}, in particular the ones which in $D=4$ are
algebraically equivalent to $\hat{o}(1,4)$ dS algebra \cite{MM_PLB814}-\cite{LMMP_PLB838}. Here we consider  relativistic models with the noncommutativity in quantum space-time and fourmomenta sectors, consistent with the covariance under the Born map \cite{Born1,Born2} and its generalizations (see \eqref{born}, \eqref{born_ext}). Large class of such models are obtained by the generalization of Yang model \cite{Yang} historically the first relativistic NC model of quantum phase space which is Born-selfdual and includes NC fourmomenta sector.

One can draw the following diagram which describes the relations between four NC relativistic models mentioned in this paper:
standard Snyder model \cite{Snyder}, its $\kappa$-deformed versions \cite{kappa-Snyder}, \cite{kappa-Snyder1}, analogously standard Yang model \cite{Yang} with its doubly $\kappa$-deformed version proposed firstly in \cite{PLB2024}.

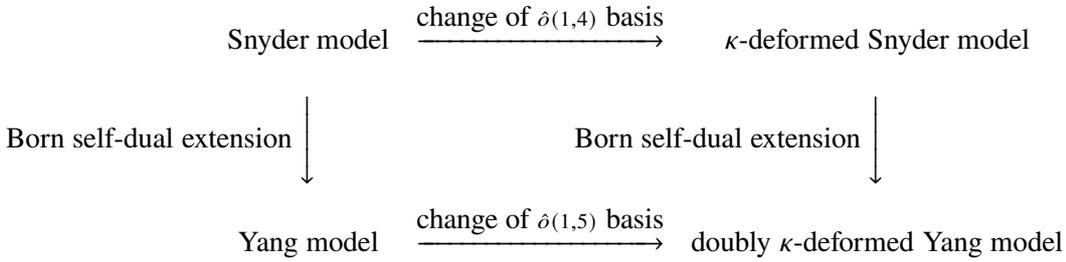
\catcode`\@=11
\newdimen\cdsep
\cdsep=3em

\def\cdstrut{\vrule height .6\cdsep width 0pt depth .4\cdsep}
\def\@cdstrut{{\advance\cdsep by 2em\cdstrut}}

\def\arrow#1#2{
  \ifx d#1
    \llap{$\scriptstyle#2$}\left\downarrow\cdstrut\right.\@cdstrut\fi
  \ifx u#1
    \llap{$\scriptstyle#2$}\left\uparrow\cdstrut\right.\@cdstrut\fi
  \ifx r#1
    \mathop{\hbox to \cdsep{\rightarrowfill}}\limits^{#2}\fi
  \ifx l#1
    \mathop{\hbox to \cdsep{\leftarrowfill}}\limits^{#2}\fi
}
\catcode`\@=12

\cdsep=3em
$$
\begin{matrix}
  \mbox{Snyder model}                  & \arrow{r}{\mbox{change of } \hat{o}(1,4) \mbox{ basis}}   & \kappa\mbox{-deformed Snyder model}                      \cr
  \arrow{d}{\mbox{Born self-dual extension}} &                      & \arrow{d}{\mbox{Born self-dual extension}} \cr
  \mbox{Yang model}                 & \arrow{r}{\mbox{change of } \hat{o}(1,5) \mbox{ basis}} & \mbox{doubly }\kappa\mbox{-deformed Yang model}       \cr
\end{matrix}
$$
We recall \footnote{See e.g. \cite{26} Sect.2.1.} that the curved NC space-times can be described by the generators of the following two algebraic cosets, where $O(n,k)$ denotes the Lie group corresponding to $\hat{o}(n,k)$ Lie algebra:
\begin{equation}
\frac{O(1,4)}{O(1,3)}\qquad\qquad\qquad\mbox{ or }\qquad\qquad\qquad\frac{O(2,3)}{O(1,3)}
\end{equation}
\begin{equation}
\mbox{Snyder-dS quantum space-time}\qquad\mbox{  }\qquad\mbox{Snyder-AdS quantum space-time}\nonumber
\end{equation}
In order to obtain the NC quantum relativistic phase spaces one can  consider four versions of the Yang model. It follows from \eqref{x_id}, \eqref{q_id} that the space-time sector is described by the fourth space axis of $\hat{o}(1,5)$ algebra and quantum fourmomenta are related with the fifth space axis. We get four types of dS-dS), (ds-AdS), (AdS-dS) and (AdS-AdS) Yang models , where at the first place we denote the type of curvature in NC space-time and in the second the type of curvature in NC fourmomenta sector. For example, in the case (dS-dS) Yang model, the NC quantum phase-space coordinates $(\hat{x}_\mu,\hat{p}_\mu)$ and internal $\hat{o}(2)$ symmetries are described by the following coset:
\begin{equation}\label{cosets}
\frac{O(1,5)}{O(1,3)}\equiv\frac{O(1,5)}{O(1,3)\otimes O(2)}\otimes O(2).
\end{equation}
If we deal with (dS-AdS) Yang model we should consider the cosets of algebra $\hat{o}(2,4)$ with signature $(-1,1,1,1,1,-1)$
\begin{equation}
\frac{O(2,4)}{O(1,3)}\equiv\frac{O(2,4)}{O(1,3)\otimes O(1,1)}\otimes O(1,1).
\end{equation}
The case (AdS-dS) is obtained by the exchange of fourth and fifth axis. The algebraic description of (AdS-AdS) Yang model requires the consideration of the coset of $O(3,3)$ (see \cite{26}, Sect.2.2) with respective NC quantum phase space described by the coset
\begin{equation}
\frac{O(3,3)}{O(1,3)}\equiv\frac{O(3,3)}{O(1,3)\otimes O(2)}\otimes O(2).
\end{equation}
Further, considering (dS-dS) version of Yang model based on $\hat{o}(1,5)$ Lie algebra, it permits to explore the following two new research directions:
\begin{itemize}
\item[i)] To consider quantum phase spaces described by internal symmetry $O(N)$ multiplets\footnote{We consider in the paper the quantum phase spaces with real NC quantum coordinates.}
\begin{equation}
(\hat{x}_\mu,\hat{q_\mu})\rightarrow (\hat{x}_{\mu;i},\hat{q}_{\mu;i})\qquad i=1,\ldots,N.
\end{equation}
For that purpose one can consider the following generalization of the algebraic basis defining the generalization of Yang models (see \cite{27}) \footnote{The algebras \eqref{o} for $N=5$ (see\cite{Chamse}) and for $N=7$ (see \cite{Roume}) have been recently used for the description of the unification of quantum gravity and the elementary particles sector.}
\begin{equation}\label{o}
\hat{o}(1,5)\rightarrow\hat{o}(1,3+2N)
\end{equation}
where $N$ describes the number of components in the internal $\hat{o}(N)$ symmetry multiplets. The relation \eqref{cosets} gets generalized as follows:
\begin{equation}\label{cosets2}
\frac{O(1,3+2N)}{O(1,3)}\equiv\frac{O(1,2+2N)}{O(1,3)\otimes O(2N)}\otimes O(2N).
\end{equation}
One can further decompose $O(2N)$ in the following way:
\begin{equation}
O(2N)\equiv \frac{O(2N)}{(O(N))^2}\otimes O(N)
\end{equation}
where 
\begin{equation}\label{coset3}
\frac{O(2N)}{(O(N))^2}\equiv 
\frac{O(2N)}{\mbox{diag}(
O(N)\otimes O(N))}.
\end{equation}
The algebra $\hat{o}(2N)$, corresponding to the Lie group $O(2N)$, contains the unbroken symmetries $\hat{o}(N)$ and the
symmetries described by the coset \eqref{coset3} should be spontaneously broken.
\item[ii)] Second idea which we promote is the consideration of algebra $\hat{o}(1,5)$ in the (dS-dS) Yang model as the Hopf algebras with nonprimitive coproducts. It would be interesting to consider e.g. in the case of Snyder model the quantum deformed $\hat{o}(1,4)$ algebras containing  quantum (with nonprimitive coproducts) Lorentz Hopf subalgebra $\hat{o}(1,3)$ \footnote{A Hopf subalgebra A of a Hopf algebra H is a Hopf algebra in itself, i.e. the multiplication, comultiplication, counit and antipode of H are restricted to A (and additionally the identity 1 of H is required to be in A). In other words: a subalgebra A of a Hopf algebra H is a Hopf subalgebra if it is a subcoalgebra of H and the antipode S maps A into A.}. Analogously in Yang model, one can consider the quantum $\hat{o}(1,5)$ Hopf algebras which do contain the nontrivial quantum Hopf subalgebras $\hat{o}(1,3)\otimes\hat{o}(2)$ .
\end{itemize}
Having such Hopf algebraic structures one can introduce new quantum versions of Snyder and Yang models described by the Hopf algebraic generalizations of \eqref{snyderx}-\eqref{snyderMM} and \eqref{yangq}, \eqref{rxrq}-\eqref{MM}. It is interesting to observe here that the complete classification of classical r-matrices generating quantum deformations of $D=4$ (A-dS) group with Lorentz generators describing its quantum subgroups has been obtained in \cite{30}. Using these results one can provide new Hopf algebraic generalizations of Snyder and Yang models.

%Finally, the most interesting would be to consider simultaneously the application of ideas presented in points i) and ii). However, firstly we should clearly understand what we gain if we extend the construction of Snyder model from classical Lie algebra $\hat{o}(1,4)$ to the suitably quantum deformed examples of quantum $\hat{o}(1,4)$ Hopf algebras. Similarly, the analogous construction of new Yang models could be achieved if we would consider quantum $\hat{o}(1,5)$ Hopf algebra with quantum-deformed $\hat{o}(1,3)\otimes\hat{o}(2)$ Hopf algebra as their quantum subalgebra.

\section*{Acknowledgements}

The authors were supported by Polish NCN grant 2022/45/B/ST2/01067.
A. Pacho\l\ 
acknowledges the support of the European Cooperation in Science and Technology COST Action CaLISTA CA21109.
% supported by COST (European Cooperation in Science and Technology). www.cost.eu.

%\section*{Appendix: From Snyder model to $\kappa$-deformation as the change of $\hat o(1,4)$ Lie algebra realizations}
%
%\ap{Change of basis}\\
%\[
%[\hat x_\mu, \hat x_\nu] = i \beta^2 M_{\mu \nu},\]
%
%\[
%[\hat p_\mu, \hat p_\nu] = i \alpha^2 M_{\mu \nu}
%\]
%
%Defining $\tilde x_\mu, \tilde p_\mu$ and $M_{\mu \nu}$ as follows (see also Kowalski-Glikman, Nowak hep-th 0304101)
%\[
%\tilde x_\mu = \hat x_\mu + \beta a_\rho M_{\mu \rho},\]
%\[
%\tilde p_\mu = \hat p_\mu + \alpha b_\rho M_{\mu \rho}
%\] one can check that they generate the same algebra 
%They satisfy \ap{to be checked}
%\begin{equation}
%[\tilde x_\mu, \tilde x_\nu] =
%i(a_\mu \tilde x_\nu - a_\nu \tilde x_\mu + \beta^2(1 + a^2) M_{\mu \nu})
%\end{equation}
%\begin{equation}
%[\tilde p_\mu, \tilde p_\nu] =
%i(b_mu \tilde p_\nu - b_\nu \tilde p_\mu + \alpha^2(1 + b^2) M_{\mu \nu})
%\end{equation}
%If $a^2 = -1$ then $\kappa$-Poincare ? algebra is equivalent to  Snyder/Yang algebra.
%
%$$***$$

%\begin{thebibliography}{99}
%\bibitem{...}
%....

%\end{thebibliography}

\end{document}